# Atomic scale characterization of deformation induced interfacial mixing in a Cu/V nanocomposite wire

X. Sauvage[1*], C. Genevois[1], G. Da Costa[1], V. Pantsyrny[2]

*1- University of Rouen, Groupe de Physique des Matériaux, CNRS (UMR 6634), Avenue de l'Université - BP 12, 76801 Saint-Etienne du Rouvray, France*

*2- Bochvar Institute of Inorganic Materials, Rogova Street 5a, VNIINM, Moscow 123060, Russia*

\* corresponding authors : xavier.sauvage@univ-rouen.fr

**Abstract**

The microstructure of a Cu/V nanocomposite wire processed by cold drawing was investigated by high resolution transmission electron microscopy and atom probe tomography. The experimental data clearly reveal some deformation induced interfacial mixing where the vanadium filaments are nanoscaled. The mixed layer is a 2nm wide vanadium gradient in the fcc Cu phase. This mechanical mixing leads to the local fragmentation and dissolution of the filaments and to the formation of vanadium super saturated solid solutions in fcc Cu.

Since the middle of the twentieth century, it has been well established that heavy drawing of two phase alloys may give rise to some very high strength materials. This is the typical case of cold drawn pearlitic steels that exhibit a mechanical strength higher than 3 GPa and that are widely used as tyre cords. Such unusual properties are attributed to the nanoscaled structure that develops during the drawing process: cementite lamellae are aligned and elongated along the wire axis and after a true strain of about three, the mean interlamellar spacing is typically less than 30nm [1]. This concept has been later extended to copper wires alloyed with Nb [2-5], Ta [6], Fe [7], Cr [8], Ag [9, 10] using various technical strategies like accumulated cold drawing [6], powder metallurgy [7] or processing of cast alloys [2, 4, 8, 10]. All these binary systems are characterized by a very low solubility of the alloying element in the fcc copper phase at room temperature and during heavy drawing, the two phase mixture is transformed into a nanoscaled filamentary structure. Such fibrous nanocomposite materials usually exhibit an excellent combination of high strength and good electrical conductivity that is attractive for many applications.

However, it is now well established that during severe plastic deformation, some unique features may occur at the nanoscale such as deformation induced phase transformation [11], interdiffusion [12, 13] or even amorphization [14], and heavily drawn two phase materials may also be affected. For instance, in cold drawn pearlitic steels, it has been demonstrated by various experimental techniques that cementite lamellae may completely decompose [15, 16]. In severely drawn Cu-Nb [5] and Cu-Ag [9] super saturated solid solutions well above the equilibrium solubility limits were observed as a result of deformation induced intermixing. In the specific case of Cu-Nb this feature was related to the partial solid state amorphization pointed out by High Resolution Transmission Electron Microscopy (HRTEM) [5, 17].

The physical mechanisms of such deformation induced intermixing have been extensively investigated for ball milled powders [18, 19]. However, the situation is very different for severely drawn wires, because the total accumulated strain and the strain rate are much lower and the material flow is continuous (which gives rise to the fibrous nanostructure). The Atom Probe Tomography (APT) technique [20, 21] is probably the most appropriate tool to characterize the nanostructure of filamentary composite and to map out diffusion gradients. This was done few years ago on Cu-Ag [9] and Cu-Nb [5] wires and the experimental data clearly show that intermixing occurs only where the filament thickness is reduced down to only few nanometers. However, these experiments raised two major limitations:

(i) The field of view of the atom probe used at that time was quite limited (typically 10x10 nm$^2$ in the cross section of the wire and 100nm along the wire axis). This is much smaller than the average length scale of the heterogeneity of the microstructure in such nanocomposite materials, thus it was extremely difficult to get a representative overview of the nanostructure. (ii) The evaporation field of Cu (30 Vnm$^{-1}$) is different from that of Nb (37 Vnm$^{-1}$) and Ag (24 Vnm$^{-1}$) [21] which could give rise to some local magnification effects and artefact in reconstructed volumes [22].

In the present study, a Cu-V nanocomposite wire was prepared by heavy drawing and the resulting nanostructure was characterized both by Scanning Transmission Electron Microscopy (STEM) and APT using a wide angle atom probe. The main advantage of this system is that Cu and V exhibit the same evaporation field (30 Vnm$^{-1}$) [21], thus it is though to be ideal for the investigation of the physical mechanisms leading to the non equilibrium intermixing occurring at the nanoscale. The Cu-V system exhibits a large positive heat of mixing ($\Delta H_{mix}$ = 5kJ/mol) and is among the so-called immiscible systems [23]. The solubility of V in fcc Cu is indeed extremely limited with a maximum of 0.8at.% at 1393K and less than 0.1at.% near room temperature [24]. On the other side of the phase diagram, the solubility of Cu in bcc V is significantly higher, up to 8 at.% at 1803K and about 5 at.% near room temperature [24]. It has been shown however that non equilibrium super saturated solid solutions can be achieved in this system by mechanical alloying [25-28], melt-spinning [29] or irradiation [30].

The Cu-V alloy was prepared by arc melting with consumable electrode from high purity electron-beam melted copper (99.99%) and vanadium (99.9%) with a molar fraction of 80% copper. The initial billet with a diameter of 100 mm was extruded down to 30 mm in diameter and then cold drawn down to 0.3mm (cumulated true strain γ = 9.2). The deformation has been performed without any intermediate heat treatment that could lead to recrystallization of V filaments. The microstructure was characterized by STEM and APT. TEM samples were prepared both in the longitudinal direction and in the cross section of the wire but only data related to these later specimens are reported here. Since the wire diameter was very small, it was first nickel coated, then sliced, mechanically grinded down to 45 μm and finally specimens were ion milled at 3keV using a GATAN PIPS 691. Observations were performed using a probe corrected JEOL ARM200F microscope operating at 200kV in the STEM mode. Image acquisition was performed on JEOL Bright Field (BF) and High Angle Annular Dark Field (HAADF) detectors.

APT samples were prepared by standard electropolishing methods (20% $H_3PO_4$ and 5% $H_2SO_4$ in methanol (vol.%) at 6V). Analyses were performed using a CAMECA Laser

Assisted Wide Angle Tomographic Atom Probe (LAWATAP) and samples were field evaporated at a temperature of 50K in UHV conditions with femto-second laser pulses (pulse duration $3.5 \cdot 10^{-10}$ s, wave length 515nm, spot size 0.1mm, energy of $2 \cdot 10^{-7}$ J, pulse repetition rate 100 kHz). Data processing was performed using the GPM 3D Data software ®.

The combination of BF and HAADF STEM images displayed in fig.1 shows the cross sectional view of the microstructure of the Cu-V nanocomposite wire. Vanadium filaments appear in dark on the HAADF image (Fig. 1(b)) because the atomic number of Cu is larger. These filaments are not homogeneously distributed within the Cu matrix, which is expected for such a composite drawn from a cast ingot [3]. They are strongly curled and this results from the non-axisymmetric elongation of the vanadium bcc filaments that are embedded in the fcc copper phase [3]. This so-called "Van Gogh sky structure" is a well known consequence of the strong single component <110> fibre texture that develops during drawing in the bcc vanadium phase [31]. This texture and also the typical <111> texture of copper were both confirmed by TEM observations of longitudinal specimens (data not shown here). The thickness of vanadium filaments is in a range of few nanometers to 30 nm. The very high ratio of thickness to width of filaments should be mentioned. Grain boundaries within the copper matrix are also clearly exhibited on the BF image (Fig. 1(a)) thanks to some diffraction contrasts. Due to the plastic flow during drawing, these grains are elongated along the wire axis, and as shown on this image, their cross sectional dimension is in a range of 50 to 200 nm. Mainly the grain size is defined by the spacing between V filaments.

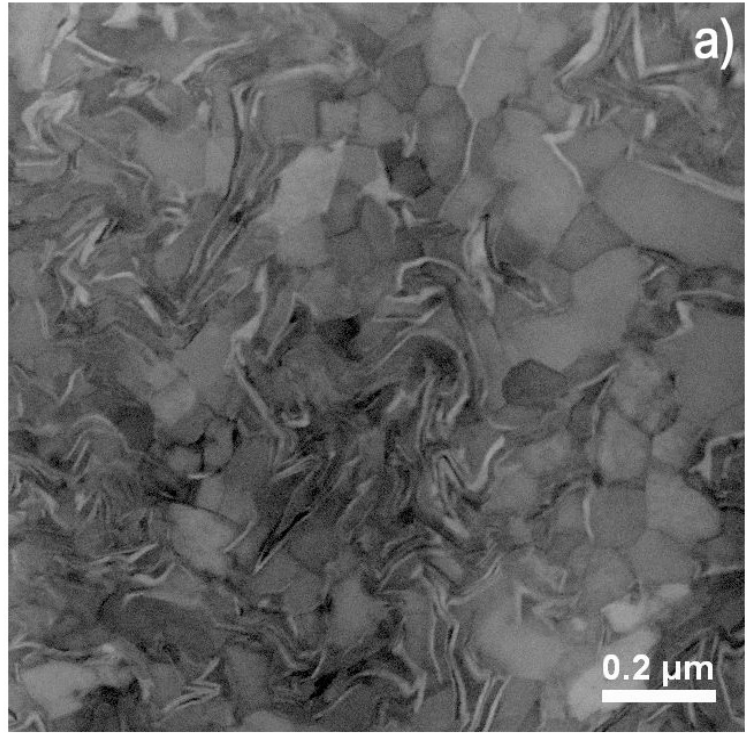

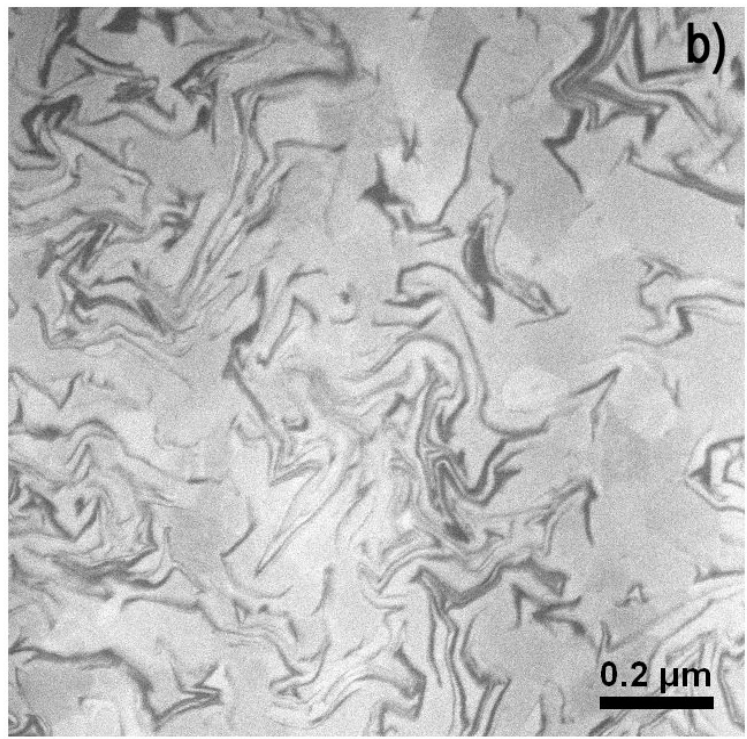

*Figure 1: STEM cross sectional view of the Cu/V nanocomposite wire drawn to a true strain of 9.2 (a) BF image where GBs are exhibited by the diffraction contrast, (b) HAADF image where V filaments are darkly imaged.*

High resolution STEM was carried out to image the nanostructure of filaments at the atomic scale. The bright field image (Fig. 2(a)) clearly exhibits atomic columns of the fcc Cu phase oriented in the <111> zone axis (right part of the image). Lattice fringes corresponding whether to (111)Cu (d=0.208 nm) or (110)V (d=0.214 nm) are also exhibited. Contrary to previous observations on Cu-Nb nanocomposite wires [5, 17] , amorphous regions were never observed in the present material. Due to the texture that develops during drawing many dislocations are stored along Cu/Nb interfaces to accommodate the large misfit between (111)Cu (d=0.208nm) and (110)Nb (d=0.233nm) [4]. In the present Cu/V composite, there is a similar texture but the misfit is one order of magnitude smaller, thus there is only few elastic energy stored at the interface and probably not enough driving force for the local solid state amorphization. On the dark field image (Fig. 2(b)), several nanoscaled vanadium filaments (dark) alternating with copper channels (bright) can be easily identified. On these images, the arrow points a crystallographic interface between the Cu matrix and a V filament. It is interesting to note that in the intensity profile computed from the HAADF image across the interface (Fig. 2(c)), a 2 nm wide gradient appears in the fcc Cu phase. This could be obviously related to some deformation induced mixing, as already reported in Cu-Nb [5] and Cu-Ag [9] nanocomposites. However, such apparent interfacial mixing could be due to some roughness of Cu/V interfaces along a direction parallel to the electron beam. To clarify this point and to get a quantitative measurement of the interdiffusion, APT analyses were carried out.

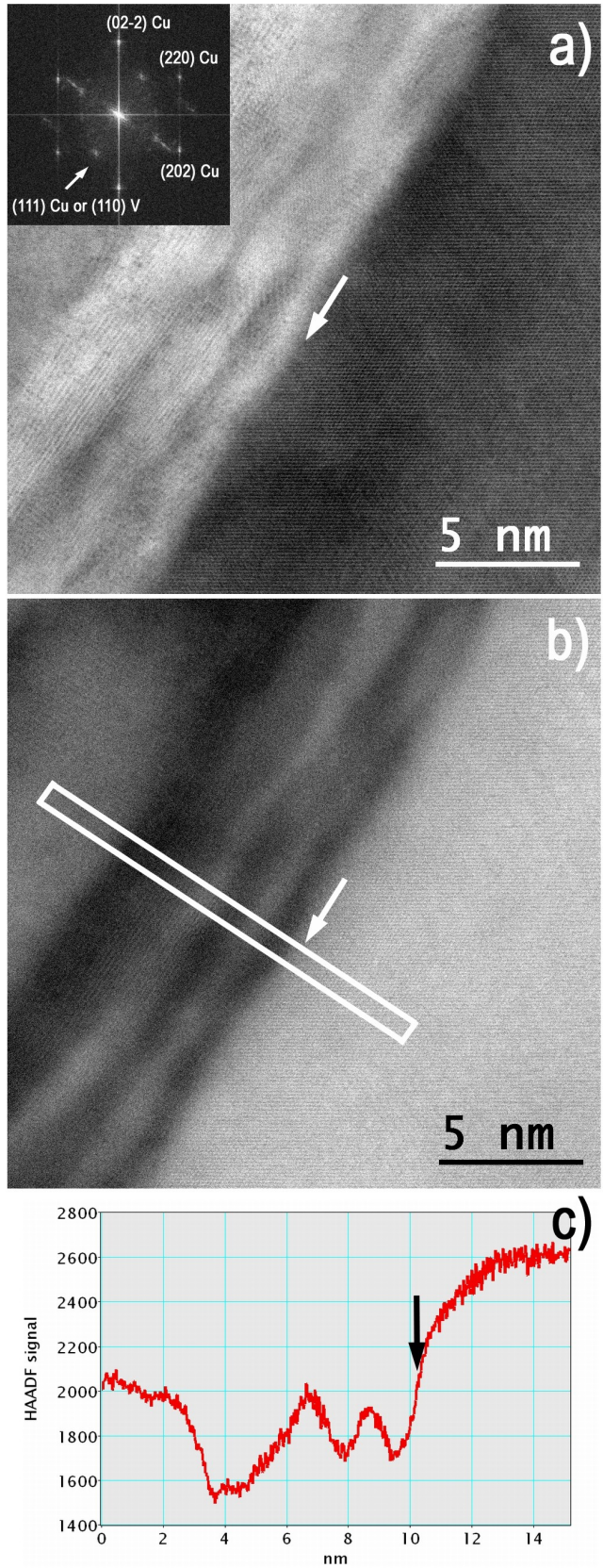

*Figure 2: HR-STEM cross sectional view of the Cu/V nanocomposite wire drawn to a true strain of 9.2 (a) BF image of the nanoscaled structrure and corresponding FFT ; (b) HAADF image showing nanoscaled V filaments (dark) in the Cu matrix (bright) ; (c) Intensity profile computed along the white box displayed in the HAADF image. The arrow located at the crystallographic interface between Cu and V, points the same position on the three images.*

The large reconstructed volume displayed in the Fig. 3(a) clearly shows vanadium nanoscaled filaments (in red) that are aligned along the wire axis. A closer view of the cross section (Fig. 3(b)) also clearly exhibits the typical curled shape as observed by STEM (Fig. 1(b)). To point out diffusion gradients, composition profiles were computed across Cu/V interfaces with a small sampling volume (section 2x4 nm² and 1nm thick). Its orientation was carefully selected to probe gradients along a direction perpendicular to interfaces. Such a profile (Fig. 3(c)) clearly shows that there are some pure Cu regions with less than 0.5at.% V in accordance with the phase diagram [24]. Some vanadium filaments are detected with a Cu concentration in a range of 3±2 to 10±4 at.% fitting also the equilibrium solubility limit [24]. However, there are obviously some non equilibrium mixed zones with compositions ranging from 50 to 65 at.%V that seem to be respectively the track of former nanoscaled Cu channels and V filaments. Such mixed regions are more clearly exhibited on the 2D concentration map (Fig. 3(d)) that displays the 2D concentration gradients of vanadium with an interpolated sampling volume of 0.3x0.3x0.3 nm³. Two vanadium filaments aligned along the wire axis are exhibited, they have a thickness in a range of only 2 to 5 nm and they obviously appear fragmented. The transition between the core of these filaments and the copper matrix is not sharp and some interfacial mixing appears over a distance of about 2nm. This is consistent with HRSTEM observations showing a similar V concentration gradient in the fcc copper phase. One should note that in the Fig. 3(d) the large copper grain located at the bottom of the image is vanadium free, while the thin copper channel located between the two vanadium filaments contains a significant proportion of vanadium atoms not homogeneously distributed. Although APT data cannot prove it, it seems realistic to think that they could be distributed around crystallographic defects like dislocations. Even more interesting is the obvious fragmentation of vanadium filaments that occurs where the surface mixed layers of the opposite sides of a filament merge.

In conclusion, the heavily deformed two phases Cu –V composite exhibits in general the typical microstructure consisting from the curled V filaments rather uniformly distributed in the Cu matrix as for much more extensively investigated Cu-Nb composite wires. In the course of deformation some interfacial mixing occurs in a thin layer of about 2nm characterized by a strong vanadium gradient in the fcc copper phase. Since V filaments are continuously elongated during the drawing process, the interface areas also increases and the global amount of V in solid solution increases. Then, when the V filament thickness is reduced down to few nanometers, mixed layers of opposite sides merge leading to its fragmentation. Contrary to Cu-Nb filamentary composite wires, there is no local amorphization probably because Cu-V interfaces contain a much smaller amount of misfit dislocations.

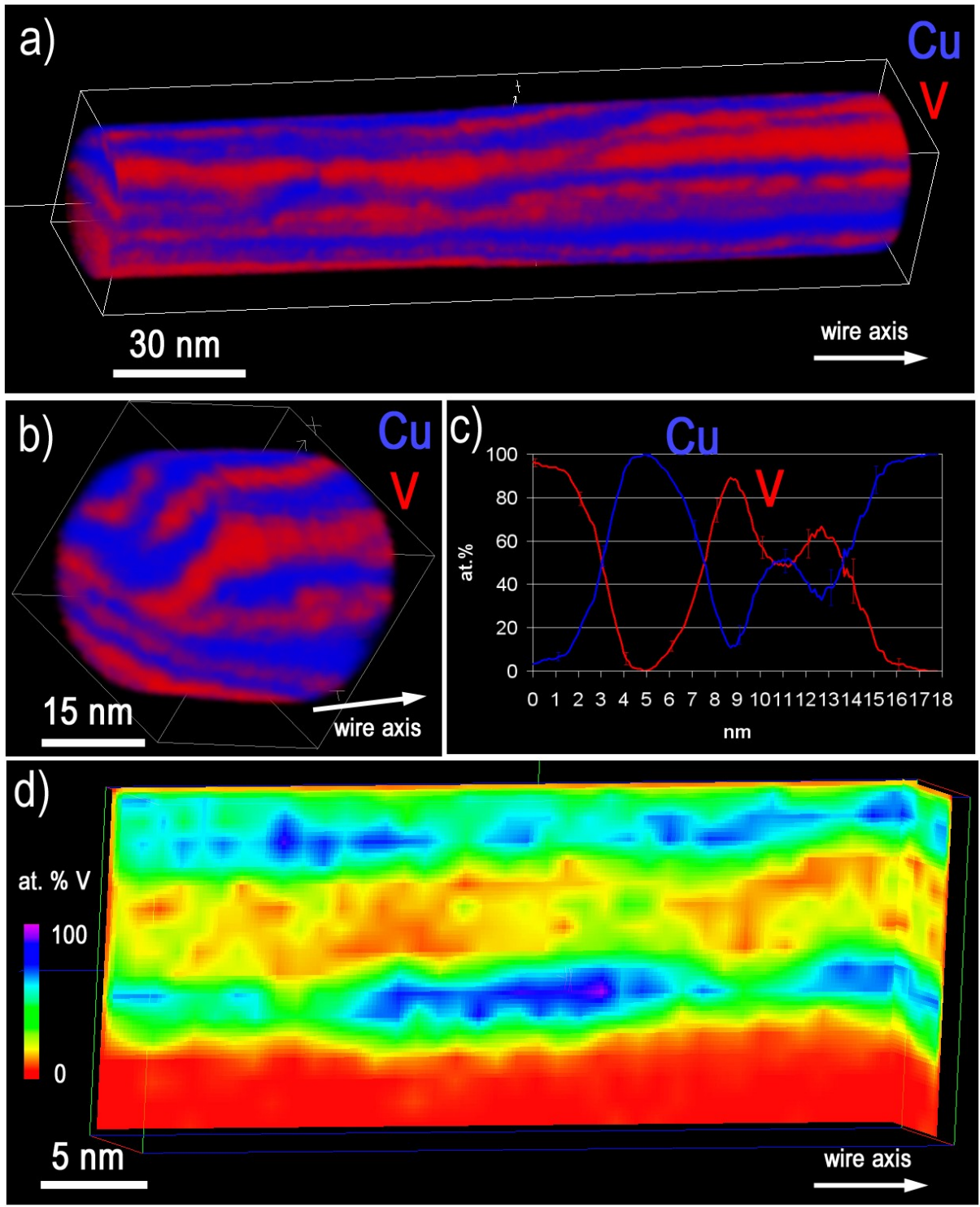

*Figure 3: APT data of the Cu/V nanocomposite wire drawn to a true strain of 9.2 (a) 3D density map computed from an APT analysed volume showing nanoscaled vanadium filaments (in red) aligned along the wire axis in the copper matrix (in blue) ; (b) Selected zone showing the typical curling of the vanadium filaments in the cross section of the wire; (c) Typical composition profile computed across several parallel Cu/V interfaces exhibiting a significant mixing (thickness of the sampling volume: 1nm); (d) Concentration 2D map showing the nanoscaled roughness and the fragmentation of vanadium filaments.*

## Acknowledgements

Dr Eiji Okunishi, Mr. Guillaume Lathus and Dr. Tetsuo Oikawa from JEOL Ltd are gratefully acknowledged for providing access to the JEOL ARM200F TEM and the image acquisition.